# `cij`: A Python code for quasiharmonic thermoelasticity

Chenxing Luo[a], Xin Deng[b], Wenzhong Wang[b,c], Gaurav Shukla[d], Zhongqing Wu[b,e,f], Renata M. Wentzcovitch[a,g,h,*]

[a] Department of Applied Physics and Applied Mathematics, Columbia University, New York, NY 10027, USA

[b] Laboratory of Seismology and Physics of Earth's Interior, School of Earth and Space, University of Science and Technology of China, Hefei, Anhui 230026, China

[c] Department of Earth Sciences, University College London, London WC1E 6BT, United Kingdom

[d] Department of Earth Sciences, Indian Institute of Science Education and Research Kolkata, Mohanpur, West Bengal, India

[e] CAS Center for Excellence in Comparative Planetology, USTC, China

[f] National Geophysical Observatory at Mengcheng, USTC, China

[g] Department of Earth and Environmental Sciences, Columbia University, New York, NY 10027, USA

[h] Lamont–Doherty Earth Observatory, Columbia University, Palisades, NY 10964, USA

[*] Corresponding author at: Lamont-Doherty Earth Observatory, Columbia University in the City of New York, 61 Route 9W, Palisades, NY 10964, USA. E-mail address: rmw2150@columbia.edu (R.M. Wentzcovitch).

**Abstract**

The Wu-Wentzcovitch semi-analytical method (SAM) is a concise and predictive formalism to calculate the high-pressure and high-temperature (high-$PT$) thermoelastic tensor (Cij) of crystalline materials. This method has been successfully applied to materials across different crystal systems in conjunction with *ab initio* calculations of static elastic coefficients and phonon frequencies. Such results have offered first-hand insights into the composition and structure of the Earth's mantle.

Here we introduce the `cij` package, a Python implementation of the SAM-Cij formalism. It enables a thermoelasticity calculation to be initiated from a single command and fully configurable from a calculation settings file to work with solids within any crystalline system. These features allow SAM-Cij calculations to work on a personal computer and to be easily integrated as a part of high-throughput workflows. Here we show the performance of this code for three minerals from different crystal systems at their relevant $PT$s: diopside (monoclinic), akimotoite (trigonal), and bridgmanite (orthorhombic).

# PROGRAM SUMMARY

*Program title:* `cij`

*Developer's repository link:* https://github.com/MineralsCloud/cij

*Licensing provisions:* GNU General Public License 3 (GPL)

*Programming language:* Python 3

*Nature of problem (approx. 50-250 words):* Experimental measurements of full elastic tensor coefficients under high-pressure and high-temperature conditions are challenging and susceptible to uncertainties. Computations of thermoelastic coefficients based on the conventional density functional theory (DFT) plus quasiharmonic approximation (QHA) or *ab initio* molecular dynamics (AIMD) methods are computationally extremely demanding, especially for materials with low symmetries because of the revaluation of free energy for strained configurations.

*Solution method (approx. 50-250 words):* Based on a semi-analytical method proposed by Wu and Wentzcovitch [1], we developed a handy code that only needs static-state elastic coefficients and phonon vibrational density of states for several equilibrium configurations at different pressure points as input to calculate the thermal elasticity. This method avoids the reevaluation of free energy for strained configurations and can be applied to all crystal systems.

*Reference:*

[1]    Z. Wu, R.M. Wentzcovitch, Phys. Rev. B 83 (2011) 184115.

# 1. Introduction

Elasticity is a fundamental property of solids that characterizes their mechanical response to external stress. Determination of elastic coefficients, especially at high pressures and temperatures (*PT*), has has wide applications in geophysics. However, despite recent methodological developments, measurement of the full elastic tensor at high *PT* has remained a challenging undertaking and susceptible to uncertainties [1,2]. Although computational methods are regularly resorted to as alternative, fully numerical *ab initio* approaches based on the density functional theory (DFT), such as DFT + QHA (quasiharmonic approximation) (e.g., Ref. [3,4]) or DFT + MD (molecular dynamics) (e.g., Ref. [5]), are computationally demanding, considering the numerous strained configurations involved, especially for crystals with low symmetry [6].

To overcome such a computational challenge, a semi-analytical method (SAM) was proposed to compute the thermoelastic tensor ($C_{ij}$) (hereafter SAM-$C_{ij}$) [7]. Compared to the traditional QHA approach, SAM-$C_{ij}$ adopts an analytical expression for the thermal part of $C_{ij}$ to circumvent the reevaluation of vibrational density of states (VDoS) for slightly strained configurations, which drastically reduces the calculation cost by at least one order of magnitude. This formalism offers an overall improved agreement with experimental measurements for high-*PT* elasticity compared to the fully numerical approach [7]. Such improvement is possibly a benefit from the imposed isotropic thermal pressure.

Other formalisms and codes have also been proposed to computationally resolve thermoelasticity, among which some are under active development [8–11]. One popular option is a quasi-static approximation (QSA), which assumes that thermal expansion accounts for the majority of thermal effects, and $C_{ij}$ vs. *T* can be approximated with $C_{ij}$ of the structure at *T* as predicted by the QHA (e.g., see Refs. [8,10–12]). This approximation can work sufficiently well up to several hundred K and is helpful to study organic molecular crystals, organic semiconductors, and metal-organic crystals. However, in geophysical applications, $C_{ij}$ usually needs to be accurately determined at several-thousand K, and this approximation has become less predictive [9]. Other formulations were also developed to calculate the thermoelastic tensor of high-symmetry materials, which usually have fewer independent Cij components [9]. These formalisms cannot avoid phonon or MD calculations for strained configurations. A similar approach [3,4] aimed at geophysical applications is nearly unfeasible for complex low-symmetry minerals of interest. In contrast, the SAM-Cij formalism has already been extensively tested for lower mantle minerals at their relevant conditions [13–20] and recently extended for low-symmetry crystals such as monoclinic and trigonal [21,22]. This method remains predictive up to the *PT* boundary of validity of the QHA (usually up to 1500-2000 K).

Here we introduce the `cij` package, a Python implementation of SAM-Cij. Unlike some other thermoelasticity calculation methods (such as Ref. [9,23]), this package is decoupled from a particular DFT software suite. As a standalone package, `cij` requires only total static energies, VDoS, and static Cij at a series of volume points as input, obtainable with most DFT software suites. One can initiate a SAM-Cij calculation from a single command and configure it within a single settings file to work with materials across different crystal systems. Since most `cij` calculations only need a few minutes to complete on a desktop-level computer, high-performance computing (HPC) setup is not imperative. Therefore, this package is easy to use on a personal computer and is ready for integration into high-throughput workflows.

This paper is organized as follows: the next section briefly reviews the SAM-Cij method; Secs. 3 and 4 describe the structure and usage of `cij`; Sec. 5 shows its application to systems of different symmetries: diopside (monoclinic), bridgmanite (orthorhombic), and akimotoite (trigonal); Sec. 6 summarizes the paper.

## 2. The SAM-Cij formalism

### 2.1. Quasiharmonic thermal elasticity

The isothermal elastic tensor elements, or elastic coefficients, $c^T_{ijkl}$, are second-order strain derivatives of the Helmholtz free energy $F$ [24]

$$c^T_{ijkl} = \frac{1}{V}\left(\frac{\partial^2 F}{\partial e_{ij} \partial e_{kl}}\right) + \frac{1}{2}P\big(2\delta_{ij}\delta_{kl} - \delta_{il}\delta_{jk} - \delta_{ik}\delta_{jl}\big), \qquad (1)$$

the QHA Helmholtz free energy $F(e,T,V)$ under strain state $e$ is [25–28]

$$F(e,V,T) = U^{\text{st}}(e,V) + \sum_{qm} \frac{1}{2}\hbar\omega_{qm}(e,V)$$
$$+ k_B T \sum_{qm} \ln\left[1 - \exp\left[-\frac{\hbar\omega_{qm}(e_{ij},V)}{k_B T}\right]\right] \quad (2)$$

where $\omega_{qm}$ are phonon frequencies of the $m$-th mode at the $q$-th wave-number. The first, second, and third terms on the r.h.s. of Eq. (2) are respectively the static total energy, $U^{\text{st}}(e,V)$, the zero-point energy, $E^{\text{zpm}}(e,V)$, and thermal excitation energy. The vibrational or phonon energy is

$$E^{\text{ph}}(e,T,V) = E^{\text{zpm}}(e,V) + E^{\text{th}}(e,T,V). \quad (2')$$

The adiabatic elastic moduli $c^S_{ijkl}$ can be converted from the isothermal one, $c^T_{ijkl}$, as [29]

$$c^S_{ijkl} = c^T_{ijkl} + \frac{T}{VC_V}\frac{\partial S^{\text{ph}}}{\partial e_{ij}}\frac{\partial S^{\text{ph}}}{\partial e_{kl}}\delta_{ij}\delta_{kl} \quad (3)$$

where $e_{ij}$ ($i,j = 1,2,3$) are infinitesimal strains and $S^{ph}(e,T,V)$ is the phonon entropy at relevant strain state.

## 2.2. Grüneisen parameters

Computing the strain derivatives in Eq. (1) requires knowledge of the variation of mode frequencies $\omega_{qm}$ w.r.t. strains $e_{ij}$, namely strain Grüneisen parameters, $\gamma^{ij}_{qm}$

$$\frac{\partial \omega_{qm}}{\omega_{qm}} = -\gamma^{ij}_{qm} e_{ij}. \quad (4)$$

SAM-Cij avoids the expensive frequency calculations for strained configurations by deriving analytical relationships between the mode average of $\gamma^{ij}_{qm}(V)$ and the mode average of volume-Grüneisen parameters, $\gamma_{qm}(V)$, which can be readily obtained from phonon calculations under hydrostatic compression. The derivation makes QHA thermal stresses hydrostatic, which is only an approximation for anisotropic materials but has the beneficial effect of producing the thermal component of Cij under this desirable stress condition. The mode-averaged strain-Grüneisen parameters necessary for longitudinal and off-diagonal Cij calculations are given by

$$\overline{\gamma^{ii}} = \frac{e_{11} + e_{22} + e_{33}}{3e_{ii}} \bar{\gamma}, \tag{5}$$

where the averages are over all $qm$ vibrational modes $\bar{\gamma} = \frac{1}{3N}\sum_{qm}\gamma_{qm}$; $e_{ii}$ ($i = 1,2,3$) are longitudinal lattice strains produced under static hydrostatic compression. They express the crystal anisotropy ignored in the isotropic approximation. Similarly, their products and derivatives are given by

$$\overline{\gamma^{ii}\gamma^{jj}} = \begin{cases} \frac{1}{5}\frac{(e_{11} + e_{22} + e_{33})^2}{e_{ii}e_{jj}}\overline{(\gamma)^2} & \text{if } i = j \\ \frac{1}{15}\frac{(e_{11} + e_{22} + e_{33})}{e_{ii}e_{jj}}\overline{(\gamma)^2} & \text{if } i \neq j \end{cases} \tag{6}$$

$$\overline{\frac{\partial \gamma^{ii}}{\partial e_{jj}}} = \begin{cases} \frac{1}{5}\frac{(e_{11} + e_{22} + e_{33})^2}{e_{ii}e_{jj}}\overline{V\frac{\partial \gamma}{\partial V}} & \text{if } i = j \\ \frac{1}{15}\frac{(e_{11} + e_{22} + e_{33})^2}{e_{ii}e_{jj}}\overline{V\frac{\partial \gamma}{\partial V}} & \text{if } i \neq j \end{cases} \tag{7}$$

where $\overline{\gamma^2} = \frac{1}{3N}\sum_{qm}\gamma_{qm}^2$, $\overline{V\frac{\partial \gamma}{\partial V}} = \frac{1}{3N}\sum_{qm}V\frac{\partial \gamma_{qm}}{\partial V}$.

### 2.3. Thermal elastic coefficients

The thermoelastic coefficients can be analytically expressed using mode-averaged strain-Grüneisen parameters, their products, and derivatives. The isothermal elastic coefficients $c_{ijkl}^T(V,T) = c_{ijkl}^{st}(V) + c_{ijkl}^{ph}(V,T)$ are the sum of static and phonon contributions (Eqs. 1-2). $c_{ijkl}^{st}(V)$ is obtained by straightforward static DFT calculations of stress vs. strain relations. In the next two sub-sections, we show how SAM-Cij evaluates $c_{ijkl}^{ph}(V,T)$ for longitudinal ($c_{iiii}^{ph}$, $i = 1,2,3$), off-diagonal ($c_{iijj}^{ph}$, $i,j = 1,2,3$, $i \neq j$) or shear ($c_{ijkl}^{ph}$, $i \neq j$ or $k \neq l$) elastic coefficients.

### 2.3.1. Longitudinal and off-diagonal elastic coefficients

For longitudinal and off-diagonal terms of the elastic tensor, Eq. (1) reduces to

$$c^T_{iijj} = \frac{1}{V}\left(\frac{\partial^2 F}{\partial e_{ii}\partial e_{jj}}\right) + P(V,T)(1-\delta_{ij}). \tag{8}$$

The phonon contribution terms are therefore

$$c^{\text{ph}}_{iijj}(V,T) = c^{\text{zpm}}_{iijj}(V) + c^{\text{th}}_{iijj}(V,T) + (1-\delta_{ij})P^{\text{ph}}(V,T), \tag{9}$$

where the phonon contribution to the pressure is $P^{\text{ph}}(V,T) = P(V,T) - P^{\text{st}}(V)$.

Combining Eqs. (2) and (4), the zero-point term becomes

$$c^{\text{zpm}}_{iijj} = \frac{\hbar}{2V}\sum_{qm}\frac{\partial^2\omega_{qm}(V)}{\partial e_{ii}\partial e_{jj}} = \frac{\hbar}{2V}\sum_{qm}\left(\gamma^{ii}_{qm}\gamma^{jj}_{qm} - \frac{\partial\gamma^{ii}_{qm}}{\partial e_{jj}} + \delta_{ij}\gamma^{ii}_{qm}\right)\omega_{qm}. \tag{10}$$

The thermal contribution $c^{\text{th}}_{iijj}$ is

$$\begin{aligned}c^{\text{th}}_{iijj}(V,T) &= \frac{k_B T}{V}\sum_{qm}\frac{\partial^2[\ln(1-e^{-Q_{qm}})]}{\partial e_{ii}\partial e_{jj}}\\ &= \frac{k_B T}{V}\sum_{qm}\Big[-Q^2_{qm}\frac{e^{Q_{qm}}}{(e^{Q_{qm}}-1)^2}\gamma^{ii}_{qm}\gamma^{jj}_{qm}\\ &\quad + \frac{Q_{qm}}{(e^{Q_{qm}}-1)}\left(\gamma^{ii}_{qm}\gamma^{jj}_{qm} - \frac{\partial\gamma^{ii}_{qm}}{\partial e_{jj}} + \gamma^{ii}_{qm}\delta_{ij}\right)\Big]\end{aligned} \tag{11}$$

where $Q_{qm} = \frac{\hbar\omega_{qm}}{kT}$. These two expressions are then simplified by using the mode-averaged values given in Eqs. (5-7) (see Ref. [7] for details). The $\frac{\partial S}{\partial e_{ii}}$ term necessary to compute Eq. (3) is given by

$$\frac{\partial S}{\partial e_{ii}} = k_B\sum_{qm}Q^2_{qm}\frac{e^{Q_{qm}}}{(e^{Q_{qm}}-1)^2}\gamma^{ii}_{qm}. \tag{12}$$

### 2.3.2. Shear elastic moduli

For the elastic tensor components that have not been addressed so far, SAM-Cij employs axis rotations to convert shear strains back to longitudinal or off-diagonal ones discussed above. To solve for $c_{ijkl}$, this SAM-Cij implementation applies a symmetric strain, $\eta^{ijkl}$, in a rotated crystal system with components given by

$$\eta^{ijkl}_{\alpha\beta} = [1 - (1-\delta_{i\alpha}\delta_{j\beta})(1-\delta_{j\alpha}\delta_{i\beta})(1-\delta_{k\alpha}\delta_{l\beta})(1-\delta_{l\alpha}\delta_{k\beta})]\xi. \tag{13}$$

9For example,

$$\eta^{2323(44)} = \begin{bmatrix} 0 & 0 & 0 \\ 0 & 0 & \xi \\ 0 & \xi & 0 \end{bmatrix} \quad \eta^{2312(46)} = \begin{bmatrix} 0 & \xi & 0 \\ \xi & 0 & \xi \\ 0 & \xi & 0 \end{bmatrix} \quad \eta^{1113(15)} = \begin{bmatrix} \xi & 0 & \xi \\ 0 & 0 & 0 \\ \xi & 0 & 0 \end{bmatrix}.$$

It is always possible to find a rotation matrix, $T$, that diagonalizes the symmetric tensor $\eta_{ij}$, i.e., a real orthonormal matrix $T$ that gives $T^{-1} \eta^{ijkl} T = \eta^{i'j'k'l'}$, where $\eta^{i'j'k'l'}$ is diagonal. The rotation matrix $T$ here is the matrix of the orthrnomal eigenvectors of $\eta^{ijkl}$, and $\eta^{i'j'k'l'}$ has corresponding eigenvalues of of $\eta^{ijkl}$ along its diagonal. Under this rotation, the invariance of strain energy gives:

$$\sum_{\alpha\beta\gamma\delta} c_{\alpha\beta\gamma\delta} \, \eta^{ijkl}_{\alpha\beta} \, \eta^{ijkl}_{\gamma\delta} = \sum_{\alpha'\beta'} c_{\alpha'\alpha'\beta'\beta'} \, \eta^{i'j'k'l'}_{\alpha'\alpha'} \, \eta^{i'j'k'l'}_{\beta'\beta'}, \tag{14}$$

where $\alpha, \beta, \gamma, \delta, \alpha', \beta' = 1,2,3$.

We note that the r.h.s. of the Eq. (14) contains longitudinal or off-diagonal terms only, which can be analytically resolved as is discussed in subsection 2.3.1 with rotated strains $T_{i'i} \, e_{ij} \, T_{jj'} = e_{i'j'}$ ($i, j, i', j' = 1,2,3$) containing negligible off-diagonal terms when $e_{11}: e_{22}: e_{33} \approx 1:1:1$.

The l.h.s. of Eq. (14), being a little more complicated, will fall into one of two scenarios:

1. For $c_{ijij}$-like terms ($i \neq j$), the only term we have on the l.h.s of Eq. (14) is the unknown term, so the equation is solvable.

2. For other $c_{ijkl}$, the l.h.s. is a combination of $c_{ijij}$-like terms ($i \neq j$) (solved in situation 1), $c_{iijj}$-like terms (solved analytically in subsection 2.3.1), and the unknown term $c_{ijkl}$. So, again, this is solvable.

A recursive algorithm is currently implemented to solve these shear terms.

## 3. The `cij` distribution

### 3.1. The distribution content

The `cij` package is written in Python 3. After decompressing the `cij.zip` zip file, one sees the Python source code in the `cij` sub-folder, input for three examples in the `examples` sub-folder, documentation in the `docs` sub-folder, and the installation script `setup.py`. The `cij` package runs on all major platforms supported by the `qha` package [27].

The Python code is organized into several modules. A description of essential modules and scripts are shown in Table 1.

### 3.2. Installation

The package can be installed with the `pip` package manager. One can directly install the package by typing in "`pip install cij`" or manually install the package by downloading the zip file "`cij.zip`" and execute "`pip install cij.zip`" at the directory of the downloaded file. The package should be ready for use after installation.

### 3.3. Program Execution

After preparing the input files (to be discussed in Sec. 4), one can navigate to the directory that contains the YAML settings file (hereafter `settings.yaml`) and execute "`cij run settings.yaml`" to perform the calculation. This command invokes the `main.py` script under the `cij/cli` directory. The flowchart in Fig. 1 will help understand the procedure of a SAM-Cij calculation.

### 3.4. Output files and plotting

A typical thermoelasticity calculation for an orthorhombic crystal with the `cij` command-line program finishes in less than one minute on a desktop computer. Each output variable specified in the `output` section of `settings.yaml` will be saved to a separate file with the same tabular $(T,V)$- or $(T,P)$-grid format as in the `qha` code [27]. The available output variables are listed in Table 2.

The cij package provides three utilities to inspect calculation results right out-of-the-box: `cij plot` converts a data table to a PNG plot; `cij extract` extracts data from the original $(T, P)$ table files and prepares data table with multiple variables at specified $T$ or $P$ for further analysis, e.g., table with $c_{ij}$'s, $K$, and $G$ vs. $P$ at 300 K; `cij extract-geotherm` extracts data and creates a data table along $PT$ of a geotherm.

### 3.5. Documentation

Detailed documentation of this program will be available online at https://mineralscloud.github.io/cij/. The source of this documentation is located in the `docs` sub-folder and can be built locally with Sphinx.

## 4. Input files

At the beginning of a calculation, the `cij run` program reads the `settings.yaml` file and two input data files that contain phonon data (hereafter `input01`) and static elasticity data (hereafter `elast.dat`). Instructions on how to prepare these files are given below (see Secs. 4.1 to 4.3).

### 4.1. The calculation settings file (`settings.yaml`)

The `settings.yaml` file is home to all calculation settings. One needs to specify calculation parameters, such as thermal EoS fitting parameters, phonon interpolation settings, input data location, and output variables to store. The available parameters and their detailed descriptions are listed in Table 3.

### 4.2. QHA input data file (`input01`)

The QHA input data file contains the static energies and phonon frequencies at various volume points. The general structure of this file is identical to the one used by the `qha` program as described in Ref. [27], but the number of formula units (`nm`) and atoms (`na`) need to be additionally appended to the end of the fourth line, after the number of volumes (`nv`), $q$-points (`nq`), and modes (`np`). The ordering of phonon mode frequencies should be matched between different volume points according to the mode symmetry to ensure proper interpolation, as described in Appendix B. The package also includes a `cij modes` utility that plots the interpolated frequencies vs. volume at a given $q$-point.

### 4.3. The static elasticity input data (`elast.dat`)

The static elasticity input data file tabulates the static elastic coefficients ($c_{ij}^{st}$, $i,j = 1$ to $6$) and axial length along three axes in Cartesian coordinates ($a_{ii}$, $i = 1,2,3$) at a series of volume points. This file format is specified in Table 4. The required elastic tensor components for each crystal system can be found in Ref. [30–32]. To compute aggregate elastic moduli, i.e., $K$ and $G$, and acoustic velocities using the Voigt-Reuss-Hill (VRH) method, unless all non-zero terms are listed, one needs to either specify the crystal system in `settings.yaml` or manually preprocess `elast.dat` with the `cij fill` utility to generate a new `elast.dat` that contains all non-zero terms. Column names in the static elastic coefficients table are invoked to compute the aggregate moduli and their ordering does not matter. The three columns of lattice parameters, `lattice_a`, `lattice_b`, and `lattice_c`, are required for all crystal systems and are not implied by the symmetry option provided in the settings file.

## 5. Examples

Here we show the high-$PT$ elasticity of three important minerals in geophysics: diopside, akimotoite, and bridgmanite. These materials' thermoelastic properties have been well-studied with SAM-Cij in Ref. [14,15,21,22]. Here, we revisit these minerals using the new `cij` package to demonstrate its reliability.

### 5.1. Diopside

Diopside, the primary host of Ca in the upper mantle, is a rock-forming pyroxene mineral with a chemical composition of $MgCaSi_2O_6$. Its structure belongs to the monoclinic crystal system, with a $C2/c$ space group. The elastic tensor of diopside contains 13 independent terms. Results shown here use the local-density approxiamation (LDA) for exchange-correlation functional [33]. Details of these DFT calculations are given in Ref. [22] and Appendix A.

Isobaric and isothermal equations of state (EoS) for diopside are shown in Figs. 2 (a, b). LDA + QHA reproduces the EoS measured at high-$P$ [34], high-$T$ [35], and high-$PT$ [36] with sufficient accuracy. The calculated thermal expansivity $\alpha$ in Fig. 2 (c) shows no obvious superlinear $T$ dependence up to 1500 K at high pressurs. However, at 0 GPa, an inflection point develops gently at ~1000 K. At higher pressures inflection points develop at approximately $T = 1170 + 37\,P$ ($T$ in K, $P$ in GPa). At pressures higher than these, the validity of the QHA might be questionable.

Fig. 3 shows the $PT$ dependence of individual elastic coefficients of diopside. The nine elastic coefficients of orthorhombic systems increase with $P$ and decrease with $T$, while the other four, $c_{15}$, $c_{25}$, $c_{35}$, and $c_{46}$ decrease with $P$ and increase with $T$. In terms of $P$-dependence, our results in Fig. 3 (a–c) are consistent with 300 K measurements from Ref. [34]. There are somewhat significant discrepancies for off-diagonal and shear $C_{ij}$ terms, but after contrasting them with the 2–3% experimental uncertainties and a significant overestimation of 24% seen in previous MD simulations [37], the discrepancies are relatively insignificant. The comparison between the calculated and measure [37] $T$-dependence of $C_{ij}$ terms made in Fig. 3 (d–f) shows good agreement up to 1000 K, except that $c_{33}$ is systematically underestimated by ~10 GPa, and $c_{23}$ has contradictory $T$-dependence against Ref. [35]. At $T > 1000$ K, measurements [35] continue to change linearly with $T$, while our results start deviating. This is likely caused by anharmonic effects and this behavior is consistent with that of the inflection point in the QHA thermal expansion coefficient shown in Fig. 2 (c).

Fig. 4 shows positive $P$ dependence, and negative $T$ dependence of the VRH averaged adiabatic bulk and shear moduli ($K_S$ and $G$) and compressional and shear velocities ($v_\mathrm{p}$ and $v_\mathrm{s}$). Our results agree well with high-$PT$ ultrasonic measurements on polycrystalline samples from Ref. [36]. Compared to Ref. [34,35], the similar deviating behavior seen in Fig. 3, which is likely caused by anharmonicity, is also observed here at $T$ beyond that of the inflection points in the QHA thermal expansion coefficient shown in Fig. 2 (c).

## 5.2. Akimotoite

$MgSiO_3$ akimotoite is a high-$P$ polymorph of pyroxene and can be stable at transition zone and uppermost lower mantle conditions in the Earth [38]. It has an ilmenite-like structure and has an $R\bar{3}$ space group with trigonal symmetry. Its strong elastic anisotropy predicted by static calculations [39] makes akimotoite an outstanding candidate for the source of large acoustic-wave anisotropy observed at the bottom of the transition zone [40,41]. A recent study [21] used SAM-Cij to investigate the $PT$ dependence of its anisotropy. Its elastic properties are fully described by 7 independent elastic tensor components. DFT details are described in Ref. [21] and Appendix A.

Fig. 5 (a, b) shows the $PVT$ EoS of akimotoite. Our LDA + QHA EoS has compatible $PT$ dependence compared to measurements at high-$P$ [42] and at high-$PT$ [43]. The systematic 1% overestimation in volume compared to measurements [42,43] is common for LDA results and can be easily reconciled with EoS corrections, if necessary [44,45]. The present LDA results are also compared to high-$PT$ generalized gradient approximation (GGA) [46] + MD EoS results from Ref. [47] after their proposed EoS correction. The comparatively superior agreement of the LDA + QHA EoS with measurements (less than 2 GPa difference) over their uncorrected GGA + MD EoS (a −6.7 GPa shift in $P$ necessary to match experimental data in Ref. [48]) justifies the choice of LDA to study thermoelasticity. This has been the case since the first elasticity calculations [3,4]. Fig. 5 (c) shows the $T$ dependence of $\alpha$ at various $P$'s. The inflection points in $\alpha$ vs. $T$ at ~ 1500–2000 K (approximated by $T = 22.5\ P + 1400$, $T$ in K, $P$ in GPa) and the $\alpha$'s superlinear dependence of $T$ beyond this boundary suggest that the QHA may be unreliable and anharmonicity might start impacting these results. Beyond this boundary, results should be treated with caution.

Fig. 6 shows the akimotoite's elastic coefficients $c_{ij}$ as functions of $P$ and $T$. Here, $c_{11}$, $c_{12}$, $c_{13}$, $c_{33}$ and $c_{44}$, increase with $P$ and decrease with $T$; $c_{14}$ and $c_{25}$ decrease with $P$ and increase with $T$. The only major conflict here is the inverted sign of $c_{14}$ in Ref. [42], which, according to Ref. [42] is caused by differences in crystal setting; once inverted, their results and ours are consistent. Other than that, our results agree well with 300 K Brillouin spectroscopy measurements in Ref. [42,48]. The smaller $c_{11}$ and $c_{44}$, and slightly larger $c_{13}$ have discrepancies comparable with experimental uncertainties. The aforementioned EoS correction [44,45], if applied, would increase $c_{11}$ and $c_{44}$, which mitigates the discrepancies further.

Fig. 7 shows the $PT$ dependence of $K_S$, $v_p$, $G$ and $v_s$. Compared to Ref. [42], our 300 K $K_S$ and $v_p$ values agree soundly, but $G$ and $v_s$ depart downwards from measurements by ~2 % when compressed. A similar feature in the $P$-dependence exists in a previous LDA calculation [39], so this is a consistent LDA prediction. The earlier reports of stiffer $K_S$ and $G$, and faster $v_p$ and $v_s$ measured at high-$PT$ with ultrasound [43] compared to both our results and Ref. [42], it is likely caused by a partially transformed sample, according to Ref. [42]. These measurements in Ref. [43] are, thus, deemed unreliable. Nevertheless, the $T$-gradient of their $K_S$ and $v_p$ are mostly aligned with ours but the that of their $G$ and $v_s$ are slightly larger than to ours.

### 5.3. Bridgmanite

$MgSiO_3$-perovskite (Mg-Pv) is the Mg-endmember of bridgmanite, the most abundant mineral in the Earth's lower mantle. The thermal $C_{ij}$ tensor of Mg-Pv calculated with SAM-Cij was used as a reference for comparison with its iron-bearing counterparts to understand the effect of alloying and iron spin-crossover [14,15,49]. Experimental determination of the thermal $C_{ij}$ tensor, especially at high-$PT$, is involved with substantial certainties [2]. High-$PT$ experimental data for pure Mg-Pv has not been published yet.

Mg-Pv has a *Pbmn* space group with orthorhombic symmetry. 9 individual elastic coefficients are required to describe its elastic properties. Calculations reported here were carried out with LDA. Calculation details are described in Ref. [14] and Appendix A.

Figs. 8 (a, b) show the $PVT$ EoS of Mg-Pv. LDA + QHA reproduces the EoS obtained with XRD-DAC measurements at high-$P$ [50–52] and high-$PT$ [51,53,54] faithfully. Compared to our LDA + QHA EoS, GGA + MD simulations [5] report a roughly less than 5% overestimated $V$ of $P, T$, due to GGA's under-binding. The reliable compression curves here allow us to proceed to calculate $C_{ij}$ at high-$PT$. Fig. 8 (c) plots $\alpha$ as a function of $T$ at several $P$'s. Emperically defined by the non-superlinear dependence of $\alpha$ on $T$, the region outlined by $\frac{\partial^2 \alpha}{\partial T^2}|_P = 0$, the black line, corresponds to the $PT$-range of QHA validity [4,55]. The $PT$ validity range generally resembles that of Ref. [55] and small variations on this boundary are caused by numerical errors in high-order $T$-derivatives, interpolation, and different choices of $PT$-grids.

Fig. 9 shows elastic coefficients $c_{ij}$ of Mg-Pv as functions of $P$ and $T$. These $c_{ij}$'s increase almost linearly with $P$ and decrease linearly with $T$. Our results agree well with those in Ref. [56] determined at ambient conditions. GGA + MD calculations from Ref. [5] disagree more with measurements. The scarcity of high-$P$ or high-$T$ measurements of elastic coefficients on pure Mg-Pv single-crystal does not allow further comparisons here, which shows why SAM-Cij results are crucial.

Fig. 10 shows the $PT$ dependence of VRH-averaged $K_S$, $G$, as well as $v_s$, and $v_p$. Derived from $C_{ij}$, these properties show uniformly positive $P$-dependence and negative $T$-dependence. Our results can be verified against 300 K and 2700 K ultrasonic measurements from Refs. [52,57] up to 100 GPa. Although Refs. [51,53] suggest a slightly larger $PT$-gradient within a narrower $PT$ range, i.e., 0–20 GPa, up to 1200 K, the inconsistencies among these measurements are either comparable or more significant than their deviation from our results. Compared to the MD simulations in Ref. [5], our SAM-Cij calculation is not only less time-consuming, but also offer much-improved consistency with experimental measurements.

## 6. Conclusion

In summary, this paper presented `cij`, an easy-to-use Python package that calculates thermal Cij, elastic moduli, and acoustic velocities for crystalline materials at high-$PT$ based on the SAM-Cij formalism. The code presented here is tested on three minerals with different crystal symmetry. Consistency between our high-$PT$ results with measurements highlights the performances of the code.

# 7. Acknowledgments

This work was supported by DOE DE-SC0019759, National Natural Science Foundation of China 41925017, and IISER-K Start-up Research Grant.

# Appendix A. DFT details

All DFT calculations were performed using the Quantum ESPRESSO [58] and the LDA exchange-correlation functional [33]. Detailed calculation parameters for these three minerals are described in the next three subsections.

## Diopside

Calculations on diopside were performed using norm-conserving pseudopotentials. For Mg, the pseudopotential was generated using the von Barth–Car's method [59,60], for Si and O the Troullier-Martins method [61] was used, for Ca, an ultrasoft pseudopotential [62] was used. The plane-wave kinetic energy cutoff was 70 Ry. Structural optimizations at 7 different $P$s were performed using variable cell-shape damped molecular dynamics (VCS-MD) [63,64] with a $2 \times 2 \times 2$ $k$-point mesh. Dynamical matrices for optimized structures were obtained using density functional perturbation theory (DFPT) [65] on a $2 \times 2 \times 2$ $q$-point mesh grid. Force constants obtained from these dynamical matrices were later interpolated to an $8 \times 8 \times 8$ mesh grid to obtain the VDoS. Strains of ±0.5% and ±1% magnitude are applied to obtain static elastic constants at each pressure.

## Akimotoite

For akimotoite, the pseudopotential of Mg was generated using the Barth–Car's method, and the pseudopotentials of O and Si were generated using Troullier-Martins' method. The plane wave cutoff energy was 70 Ry. The structure of akimotoite were optimized using the VCS-MD. with a $4 \times 4 \times 4$ $k$-point mesh at 8 different $P$s. The dynamical matrices for akimotoite were calculated using DFPT [65] on a $2 \times 2 \times 2$ $q$-point mesh and then extrapolated to a denser $4 \times 4 \times 4$ $q$-mesh to obtain VDoS. Strains of ±0.5% and ±1% magnitude are applied to obtain static elastic constants at each pressure.

## Bridgmanite

Calculations on Mg-Pv were performed on a 40-atom supercell. Ultrasoft [61] pseudopotentials were used for Al, Fe, Si, and O. A norm-conserving pseudopotential generated with von Barth-Car's method was used for Mg. The electronic states were sampled on a $2 \times 2 \times 2$ $k$-point grid with a plane-wave kinetic energy cutoff of 40 Ry, respectively. The Mg-Pv structure was optimized at 10–12 relevant $P$ points with VCS-MD. Dynamical matrices were calculated using DFPT on a 2 × 2 × 2 $q$-point grid for these structures and then interpolated to an 8 × 8 × 8 $q$-point grid to obtain VDoS. Strains of ±1% magnitude are applied to obtain static elastic constants at each pressure.

# Appendix B. Matching phonon modes for different pressure points

In order to calculate numerically mode-Grüneisen parameters, $\gamma_{qm}$, with the expression $\gamma_{qm} = -\partial \ln \omega_{qm} / \partial \ln V$, we need to interpolate $\omega_{qm}$ vs. $V$. Accurate determination of this numerical derivative for each mode would require us to identify the "same" mode at different volume points and order them accordingly in the input files. Because in outputs of DFT software, such as Quantum ESPRESSO, the mode frequencies for each volume and at each $q$-point are usually ordered by their magnitude, we will have to sort the phonon modes to align them to establish continuity between volume points. It would be impractical to sort these modes manually for complex crystals and calculations with many $q$-points, so we provide an automatic solution. There has been some effort to address the mode frequency continuity between volume points [66]. Between different $q$-points within the Brillouin zone [67], a popular, practical, and most straightforward way is to sort the phonon modes based on the eigenvectors, which is also how we implement in our code.

Assume there is no degeneracy. At reciprocal coordinate $\boldsymbol{q}$, $\omega_m^2$ and $e_{qm}$ are the $m$-th eigenvalue and eigenvector of the dynamical matrix $D(\boldsymbol{q})$ [68],

$$D(\boldsymbol{q})\, e_{qm} = \omega_{qm}^2\, e_{qm} \qquad (B.1)$$

The orthonormality of the set of normal modes $e_{qm}$ is given by

$$e_{qm}^\dagger(V)\, e_{qm'}(V + \mathrm{d}V) = \delta_{mm'} \qquad (B.2)$$

Here $e_{qm}$, $e_{qm'}$ as well as their product $A = e_{qm}^\dagger(V)\, e_{qm'}(V + \mathrm{d}V)$ are all $3N \times 3N$ matrices, and $\delta'_{mm}$ is a unitary matrix.

For crystal under two close compression states $V$ and $V + \mathrm{d}V$, it is sufficient to say the eigenvectors are "nearly orthonormal", a condition that can be expressed as

$$e_{qm}^\dagger(V)\, e_{qm'}(V + \mathrm{d}V) = \delta_{mm'} + O(\mathrm{d}V) \qquad (B.3)$$

where $O(dV)$ a $3N \times 3N$ matrix with elements is sufficiently small.

Our code includes a function `cij.misc.evec_sort` to match phonon modes at close-by volumes. The function takes an array of $3N$ frequencies $\omega_m^{(i)}$ ($m = 1$ to $3N$) for volume point $V^{(i)}$, their orthonormal eigenvectors ($3N \times 3N$ matrix, $e_m^{(i)}$) in corresponding order, and orthonormal eigenvectors ($3N \times 3N$ matrix, $e_m^{(j)}$) from another volume $V^{(j)}$. The algorithm first calculates absolute inner product $A = \text{abs}[(e_m^{(i)})^\dagger e_m^{(j)}]$. Then it identifies row index $m$ and column index $m'$ of the element with maximum values in $A$ (i.e., $mm' = \text{argmax}(A)$) to match $\omega_m$ with $\omega_{m'}$. The algorithm then sets the entire $m$-th row and $m'$-th column to zero and looks for other pairs of corresponding modes. This process is looped for $3N$ times until all modes are matched for the $q$-th $q$-point in volume points $V^{(i)}$ and $V^{(j)}$.

The algorithm works well with phonons calculated with primitive cells, where there are no or very few degenerate modes but might not be as work well in non-primitive cells and with large degeneracies [67]. We also note that Ref. [69] has proposed an alternative way to obtain Grüneisen parameters, but their method requires calculations to be performed on large supercells, this might not go well with DFT calculations, which is usually how we prepare input for this software.

In the three examples we enclosed with this code, we found that the ordering of mode frequencies does not affect the final result significantly. This is probably because the average $\bar{\gamma}$, $\overline{\gamma^2}$, and $\overline{Vd\gamma/dV}$ are used in the calculation, and not many phonon crossings occur in these examples. But this function is included in the case calculations are carried out for materials with phonon modes with abnormal volume dependence, which results in more crossings between modes.

Another function `cij.misc.evec_load` is also supplied to help users parse and load eigenvectors from Quantum ESPRESSO's `matdyn.x` output.

# Figures

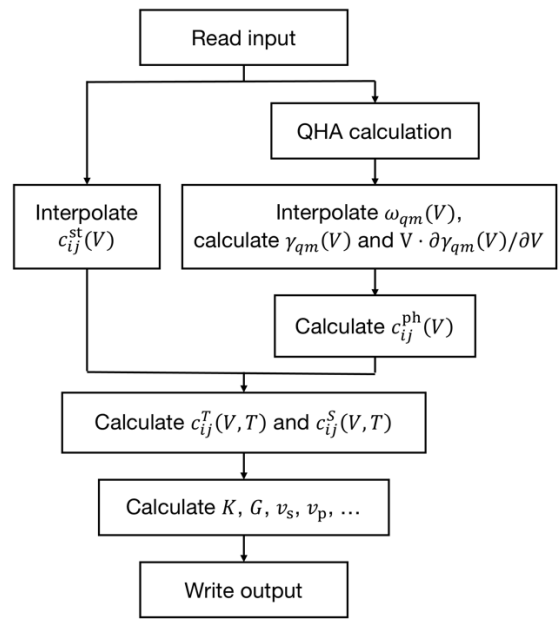

Figure 1. The flowchart for thermoelasticity calculations with SAM-Cij.

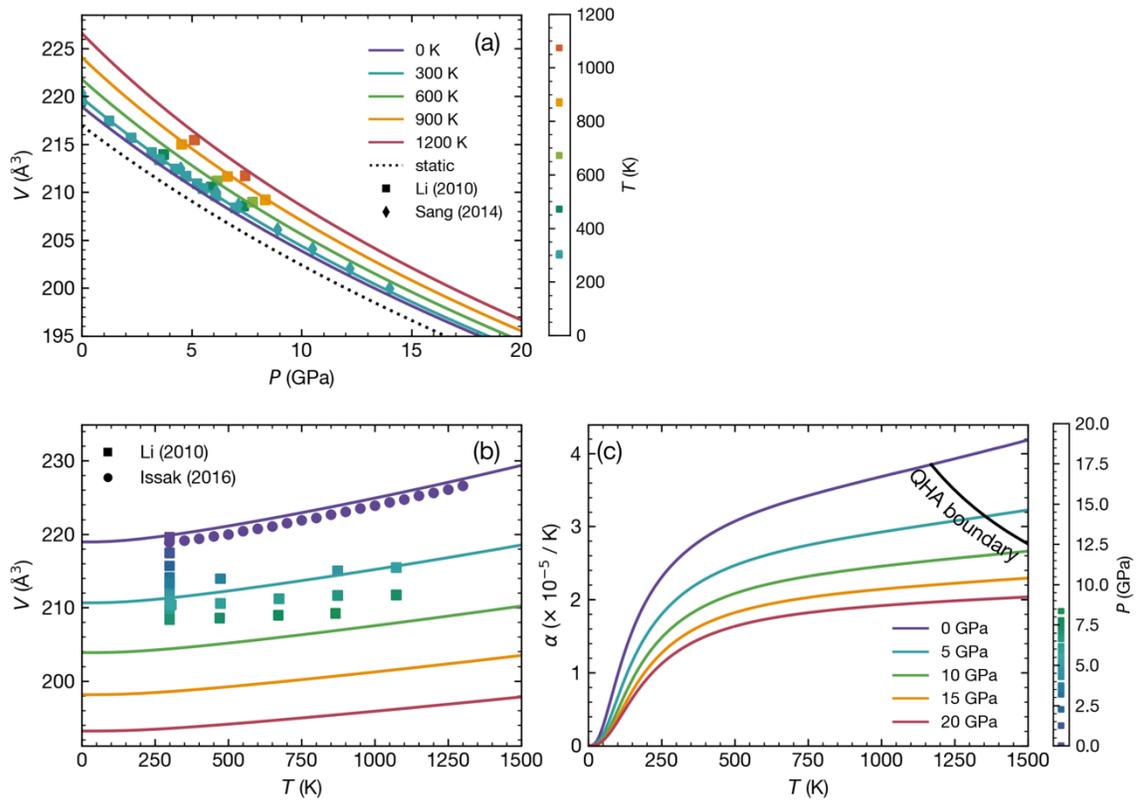

Figure 2. (a, b) Unit-cell volume (2 f.u.) of diopside (a) vs. $P$ at various $T$s, and (b) vs. $T$ at various $P$s. Teal diamonds in (a) correspond to measurements reported in Ref. [34] at 300 K, solid purple circles in (b) correspond to measurements reported in Ref. [35] at 0 GPa. Colored squares in (a, b) correspond to measurements reported in Ref. [36]. In (a) $T$ and (b) $P$ are represented by colors in the color-bars. (c) Thermal expansivity of diopside vs. $T$ at various $P$s.

24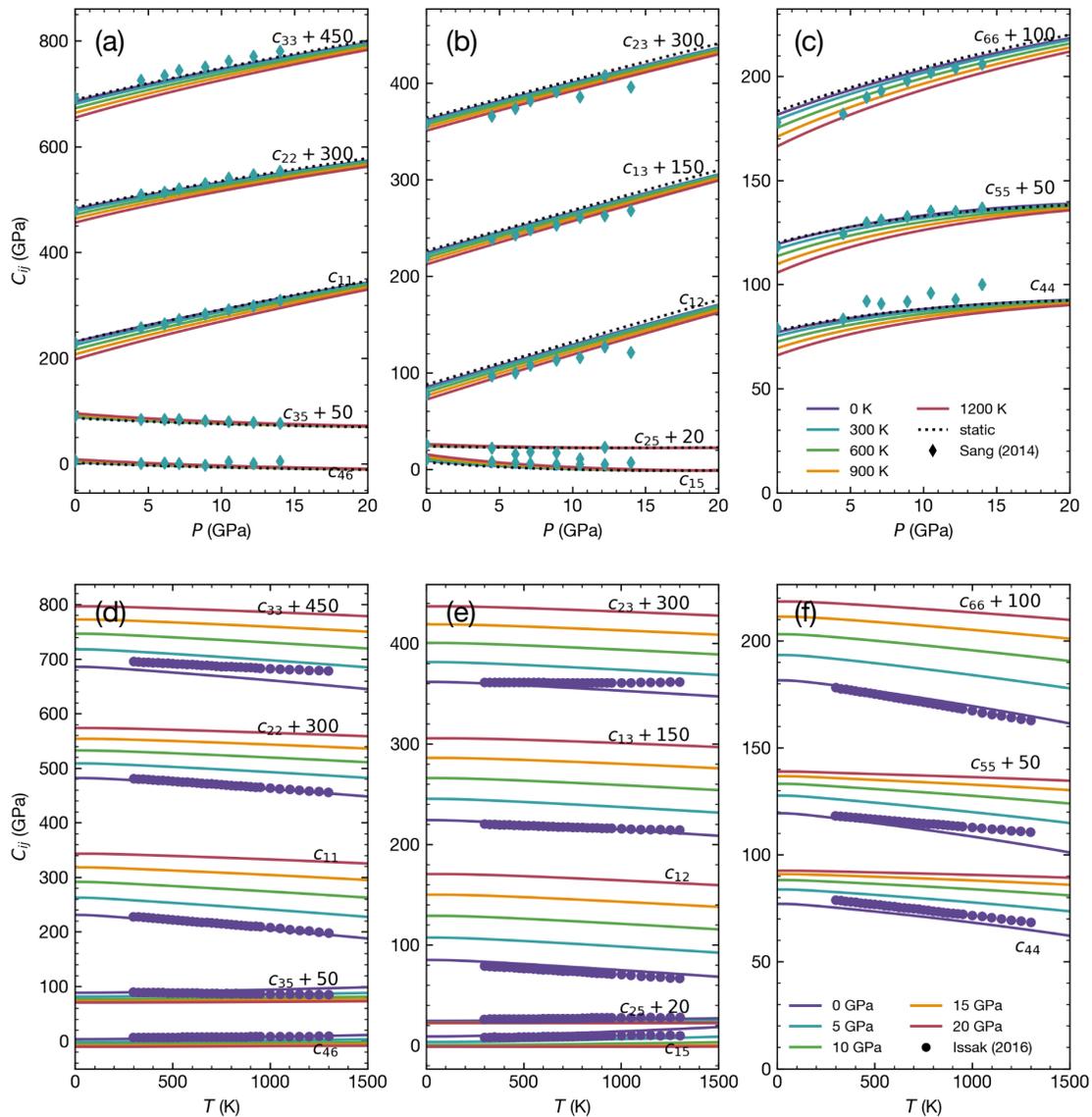

Figure 3. Elastic tensor components ($c_{ij}$) of diopside vs. (a–c) $P$ at various $T$, and (d–f) vs. $T$s at various $P$s. Teal diamonds in (a–c) correspond to measurements reported in Ref. [34] at 300 K, solid purple circles in (d–f) correspond to measurements reported in Ref. [35] at 0 GPa.

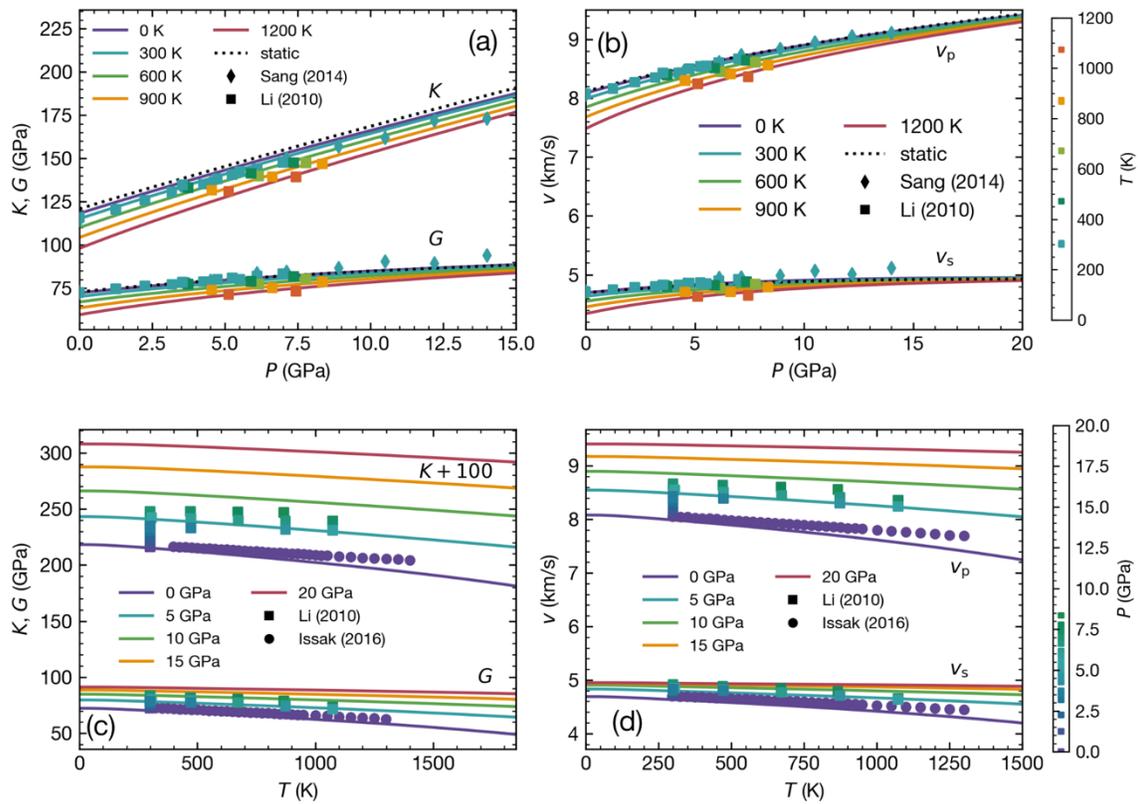

Figure 4. (a, c) Elastic moduli and (c, d) acoustic velocities of diopside vs. (a, b) $P$ and (c, d) $T$. Teal diamonds in (a, b) correspond to measurements reported in Ref. [34] at 300 K, solid purple circles in (c, d) correspond to measurements in Ref. [35] at 0 GPa. Colored squares correspond to calculations in Ref. [36] at (a, b) $T$ and (c, d) $P$ represented by colors in the color-bars.

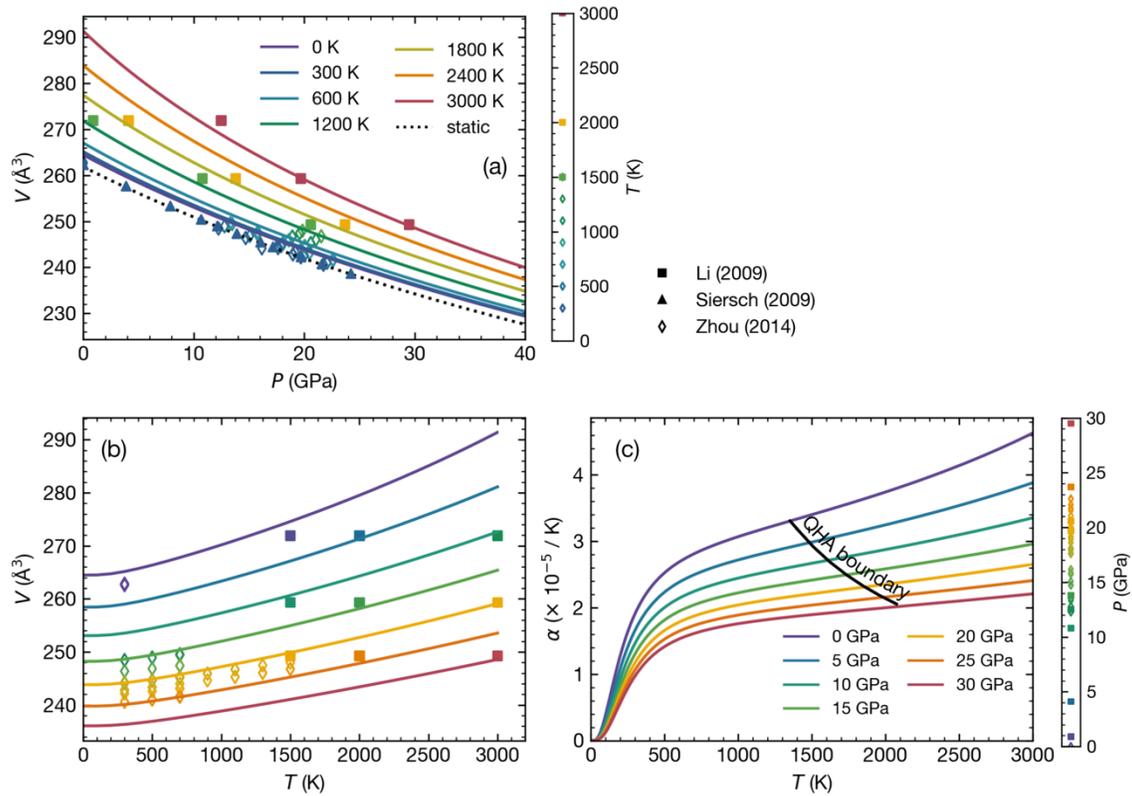

Figure 5. (a, b) Unit-cell volume (6 f.u.) of akimotoite (a) vs. $P$ at various $T$s, and (b) vs. $T$ at various $P$s. Dark blue triangles in (a) correspond to measurements reported in Ref. [42] at 300 K. Colored symbols in (a, b) correspond to calculations reported in Ref. [47] and measurements in Ref. [43] at (a) $T$s and (b) $P$s represented by colors in the color-bars. (c) Thermal expansivity of akimotoite vs. $T$ at various $P$s.

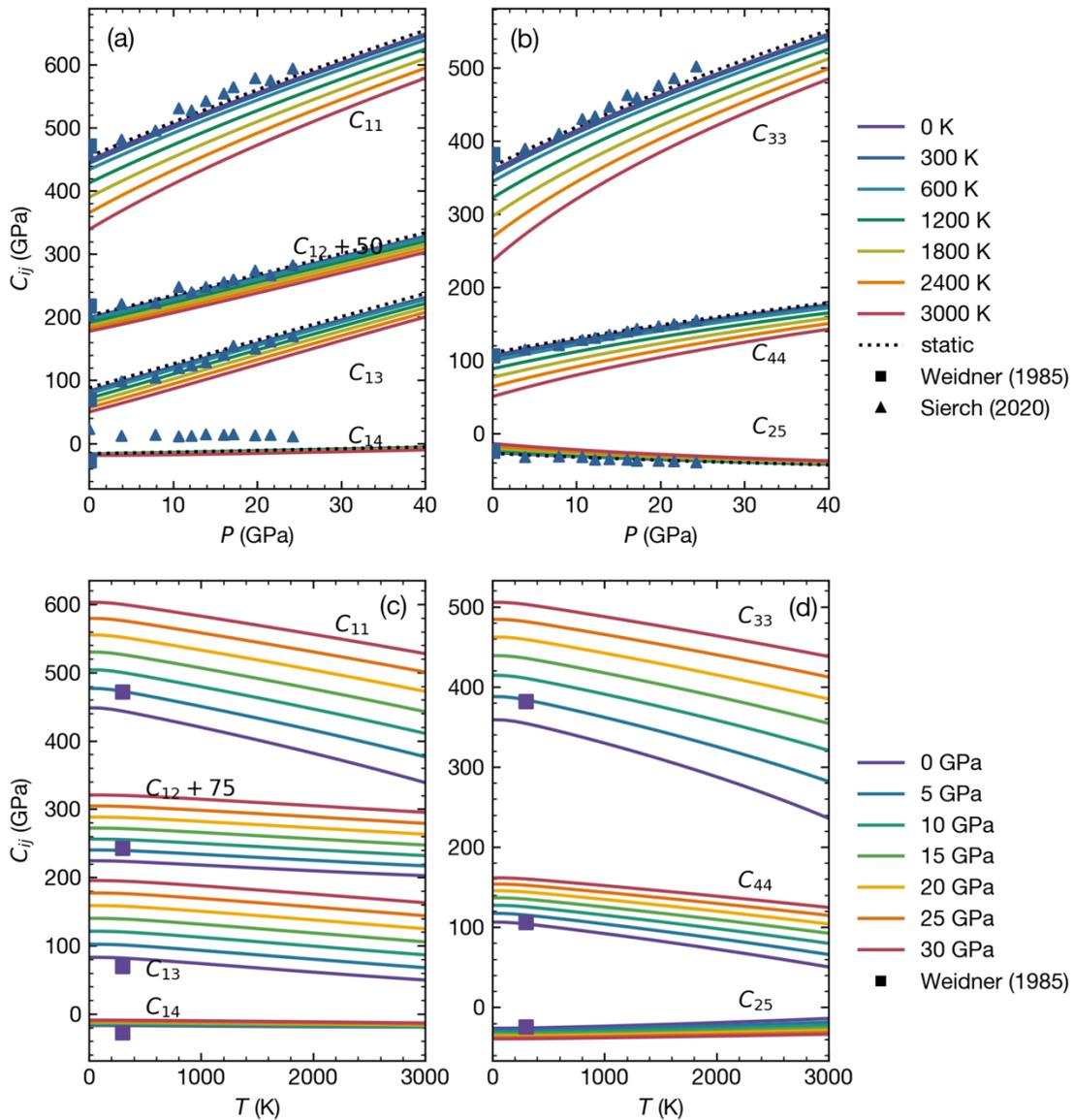

Figure 6. The $c_{ij}$'s of akimotoite vs. (a, b) $P$ at various $T$s and (c, d) vs. $T$ at various $P$s. Blue squares and triangles in (a, b) correspond to measurements in Ref. [48] and Ref. [42] at 300 K, purple squares in (c, d) correspond to measurements in Ref. [48] at 0 GPa.

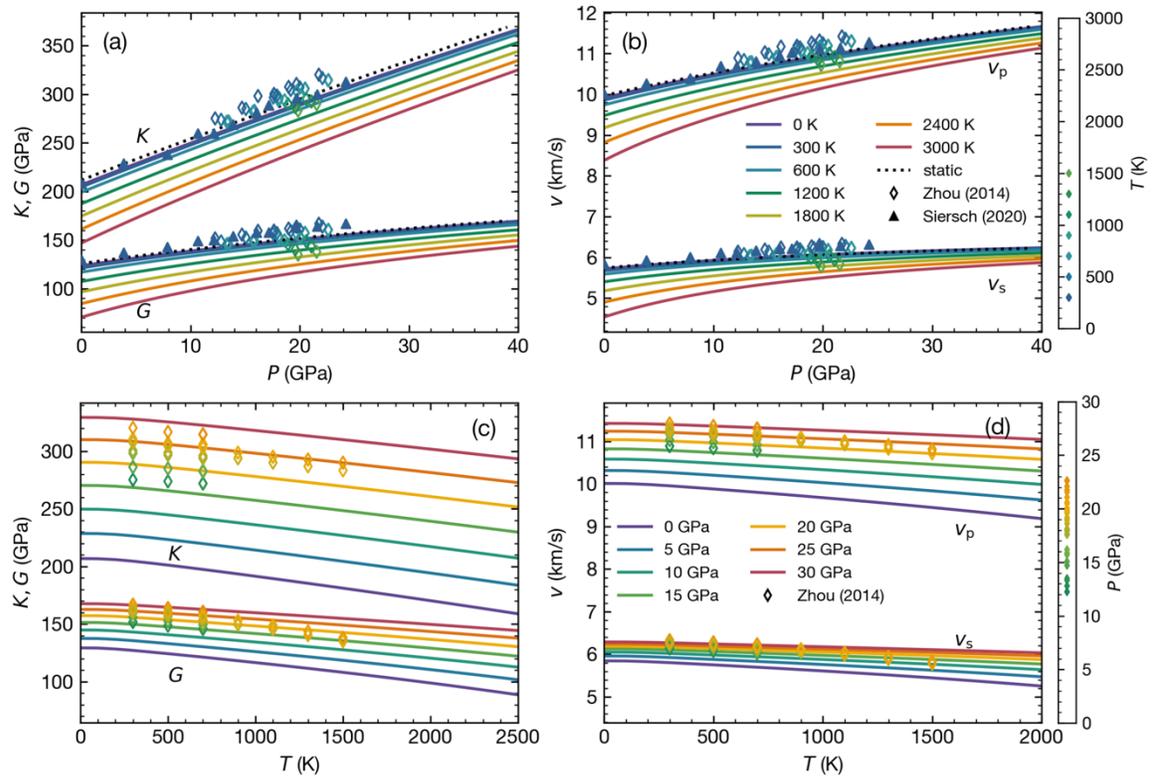

Figure 7. (a, c) Elastic moduli and (c, d) acoustic velocities of akimotoite vs. (a, b) $P$ and (c, d) $T$. Dark blue triangles in (a) correspond to measurements reported in Ref. [42] at 300 K. Colored symbols correspond to measurements reported in Ref. [43] at (a) $T$s and (b) $P$s represented by colors in the color-bars.

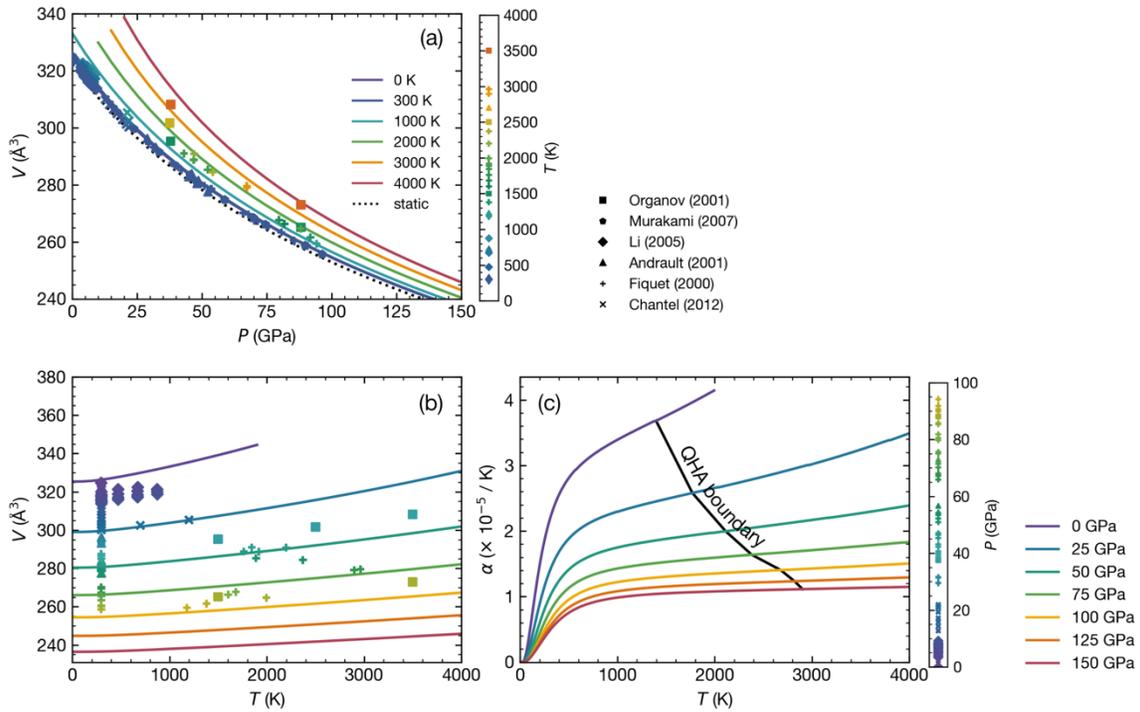

Figure 8. (a, b) Unit-cell volume (16 f.u.) of Mg-Pv (a) vs. $P$ at various $T$s, and (b) vs. $T$ at various $P$s. Dark blue pentagons and triangles in (a, b) correspond to measurements at 300 K reported in Refs. [50,52]. Colored squares, diamonds, pluses, and crosses correspond to high $PT$ GGA + MD calculations reported in Ref. [5] and measurements in Ref. [51,53]. Their $T$s (a) and $P$s (b) are represented by colors in color-bars.

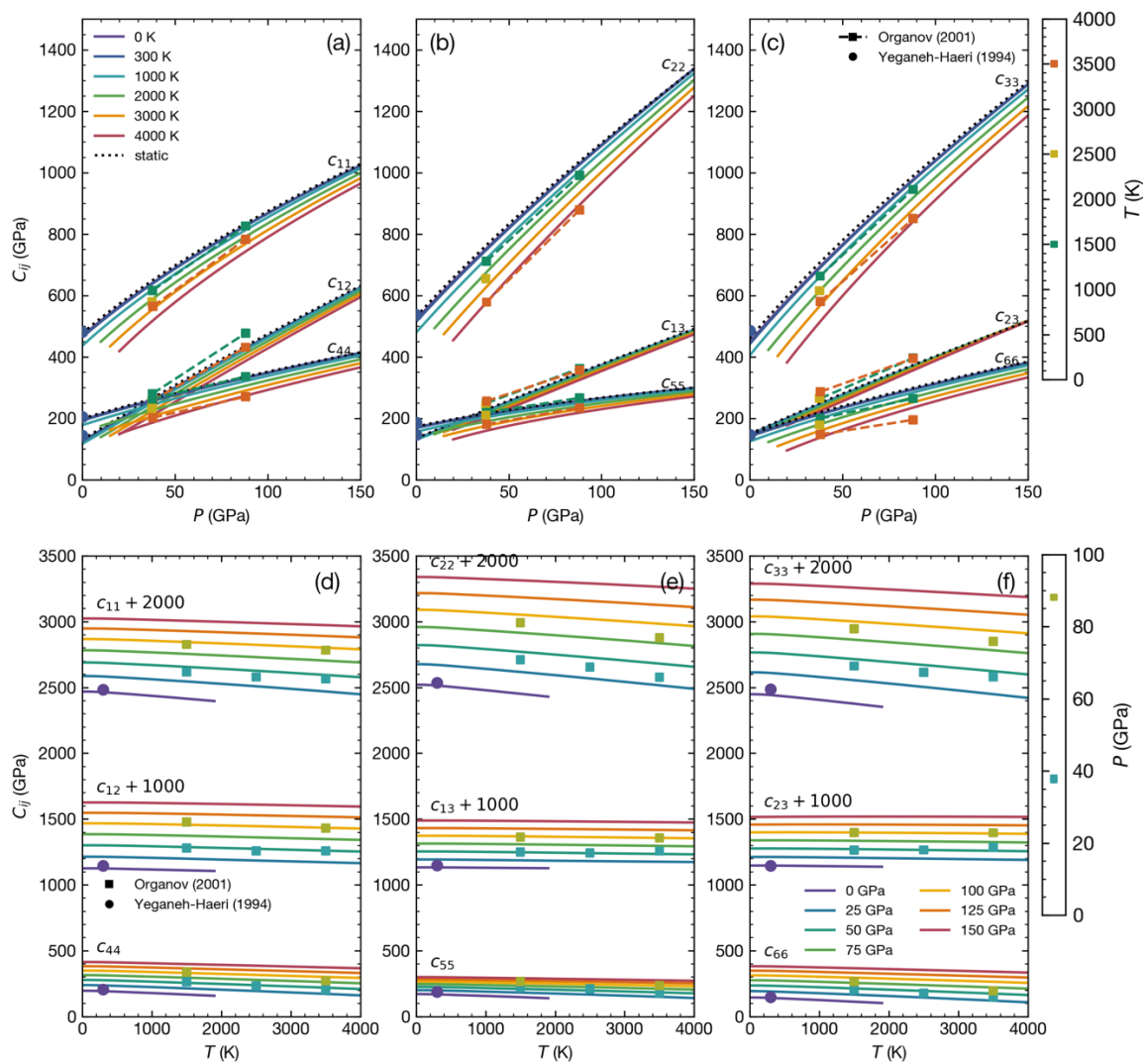

Figure 9. The $c_{ij}$'s of Mg-Pv vs. (a–c) $P$ at various $T$s, and (d–f) vs. $T$ at various $P$s. Dark blue (a–c) and purple (d–f) circles correspond to 300 K and 0 GPa measurements reported in Ref. [56]. Colored squares correspond to GGA + MD results reported in Ref. [5]. Their $T$s (a–c) and $P$s (d–f) are represented by colors in colorbars.

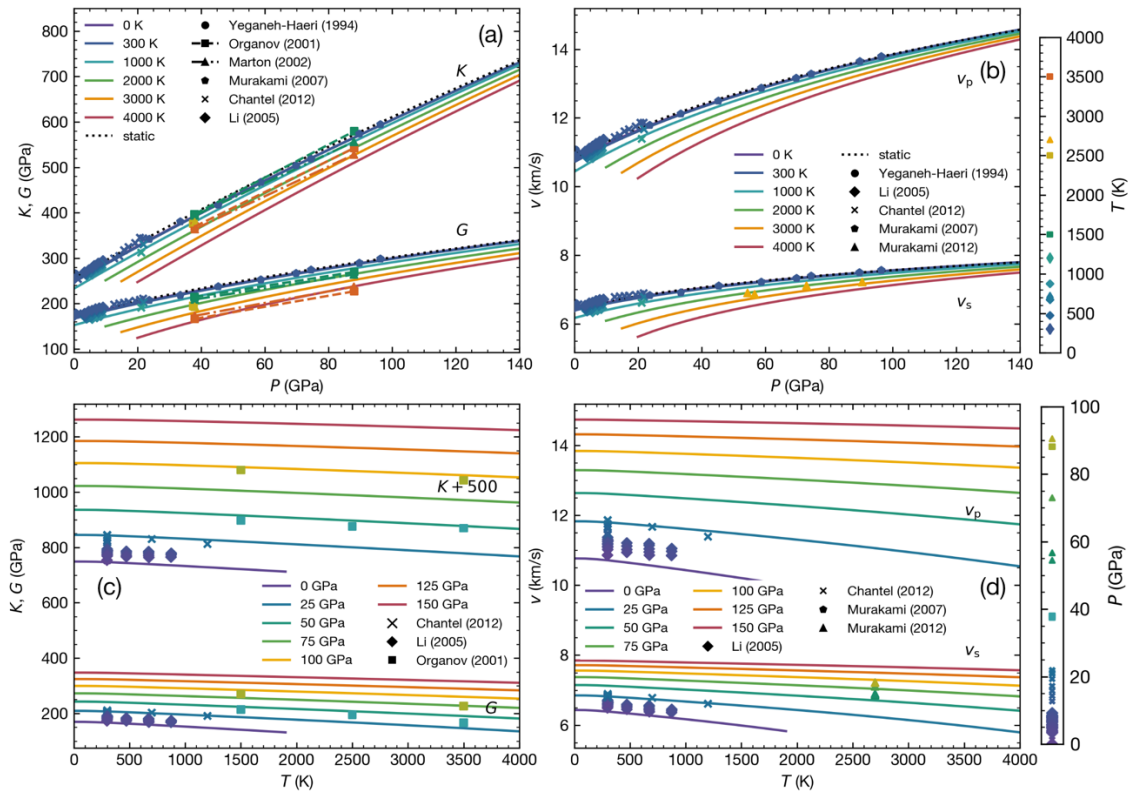

Figure 10. (a, c) Elastic moduli and (c, d) acoustic velocities of Mg-Pv vs. (a, b) $P$ and (c, d) $T$. Dark blue (a–c) and purple (d–f) circles in (a, b) correspond to 0 GPa and 300 K measurements reported in Ref. [56]; dark blue pentagons in (a) correspond to 300 K measurements [52]. Colored squares, triangles, crosses, and diamonds correspond to high $PT$ GGA + MD results reported in Ref. [5] and measurements reported by [51,53,57]. Their $T$s (a–c) and $P$s (d–f) are represented by colors in color-bars.

# Tables

Table 1. List of modules and command-line utilities in the **cij** distribution

| Module | Description |
| --- | --- |
| **cij.core** | Core functionalities<br>• **calculator** – The calculator that controls the workflow.<br>• **mode_gamma** – Interpolate phonon frequencies and calculate mode Grüneisen parameters.<br>• **phonon_contribution** – Calculate $c_{ij}^{\text{ph}}$.<br>• **full_modulus** – Interpolate $c_{ij}^{\text{st}}$ vs. $V$, and calculate $c_{ij}^{S}$ and $c_{ij}^{T}$.<br>• **tasks** – Handles the ordering of $c_{ij}^{\text{ph}}$ calculation. |
| **cij.util** | Methods used in the main program<br>• **voigt** – Convert between Voigt ($c_{ij}$) and regular ($c_{ijkl}$) notations of elastic coefficients.<br>• **units** – Handle unit conversion. |
| **cij.io** | Input output functions and classes. |
| **cij.plot** | Plotting functionalities. |
| **cij.cli** | Command-line programs<br>• **cij run** (main.py) – Perform a SAM-Cij calculation.<br>• **cij run-static** (static.py) – Calculate static elastic properties.<br>• **cij extract** (extract.py) – Extract calculation results for a specific $T$ or $P$ to a table.<br>• **cij extract-geotherm** (geotherm.py) – Extract calculation results along geotherm $PT$ (given as input) to a table.<br>• **cij plot** (plot.py) – Convert output data table to PNG plot.<br>• **cij modes** (modes.py) – Plot phonon frequency interpolation results.<br>• **cij fill** (fill.py) – Fill all the non-zero terms for elastic coefficients given the constraint of a crystal system. |
| **cij.data** | Data distributed with the program, e.g., the relationship between $c_{ij}$'s for different crystal systems, the naming scheme for output files, etc. |
| **cij.misc** | Miscellaneous functionalities that are not used in the main program, e.g., methods that facilitate the preparation of input files. |

Table 2. Relavent keywords and their options in `settings.yaml`.

| **Under "qha.settings"** | | |
|---|---|---|
| DELTA_P_SAMPLE | number | Pressure-sampling interval, used for output. The default value is 1 GPa. |
| DELTA_P | number | The interval between two nearest pressures on the grid, in GPa. The default value is 0 GPa. |
| P_MIN | number | The minimum pressure in GPa. |
| NTV | integer | Number of volumes (or equivalently, pressures) on the grid. |
| T_MIN | number | The minimum temperature, in Kelvin. The default value is 0 K. |
| DT | number | The interval between two nearest temperatures on the grid, in Kelvin. |
| NT | integer | The number of temperatures on the grid. The default value is 16. |
| **Under "elast.settings"** | | |
| symmetry.system | string | The crystal system used, one of: `triclinic`, `monoclinic`, `hexagonal`, `trigonal6`, `trigonal7`, `orthorhombic`, `tetragonal6`, `tetragonal7`, `cubic`, the default value is `trigonal`. |
| mode_gamma.interpolator | string | The method to interpolate phonon frequencies vs. volume, one of: `spline`, `lsq_poly`, `lagrange`, `krogh`, `pchip`, `hermite`, `akima`. The default value is `lsq_poly`. |
| mode_gamma.order | integer | The order of phonon frequencies spline interpolation. The default value is 3. |
| **Under "qha" and "elast"** | | |
| input | string | The location of the input files. The default value is `elast.dat`. |

Table 3. The output variable, their keyword in settings output section, default output unit, and their output file naming conventions.

| Property | Keyword | Unit | Output file naming convention ($i, j = 1$ to 6; base = tp or tv) |
|---|---|---|---|
| Adiabatic elastic modulus | cij_s | GPa | c{ij}s_{base}_gpa.txt |
| Isothermal elastic modulus | cij_t | GPa | c{ij}t_{base}_gpa.txt |
| Voigt average of bulk modulus | bm_V | GPa | bm_V_{base}_gpa.txt |
| Reuss average of bulk modulus | bm_R | GPa | bm_R_{base}_gpa.txt |
| Voigt-Reuss-Hill average of bulk modulus | bm_VRH | GPa | bm_VRH_{base}_gpa.txt |
| Voigt average of shear modulus | G_V | GPa | G_V_{base}_gpa.txt |
| Reuss average of shear modulus | G_R | GPa | G_R_{base}_gpa.txt |
| Voigt-Reuss-Hill average of shear modulus | G_VRH | GPa | G_VRH_{base}_gpa.txt |
| Shear acoustic wave velocities | v_s | km/s | v_s_{base}_km_s.txt |
| Compressive acoustic wave velocities | v_p | km/s | v_p_{base}_km_s.txt |
| Pressure vs. volume | p | GPa | v_tp_gpa.txt |
| Volume vs. pressure and temperature | V | $Å^3$ | p_tv_ang3.txt |

Table 4. The structure of static elastic coefficients input data file (`elast.dat`)

| Structure of input data | Description |
|---|---|
| **# comment line** | |
| $V_0$ N $m_{cell}$ | The calibration volume $V_0$ for static elastic moduli interpolation; the number of volumes included in this data file $N$; total cell mass, $m_{cell}$, in amu for calculation of acoustic wave velocities calculations. |
| "`v c11 c22 c33 …`" | Column names of the input data. The output elastic moduli are named after this list, the ordering of columns does not matter. |
| $V_1$ $c_{11}[V_1]$ $c_{22}[V_1]$ … | The first volume and elastic moduli at this volume; the order corresponds to the column names specified above. |
| $V_2$ $c_{11}[V_2]$ $c_{22}[V_2]$ … | Similarly organized data for subsequent volumes. |
| … | |
| $V_N$ $c_{11}[V_N]$ $c_{22}[V_N]$ … | |
| "`lattice_a lattice_b lattice_c`" | Column names for the axial lengths table. |
| $a_{11}[V_1]$ $a_{22}[V_1]$ $a_{33}[V_1]$ | The axial length along three axes of Cartesian coordinates for the first volume. |
| $a_{11}[V_2]$ $a_{22}[V_2]$ $a_{33}[V_2]$ | The axial length for subsequent volumes. |
| … | |
| $a_{11}[V_N]$ $a_{22}[V_N]$ $a_{33}[V_N]$ | |